\documentclass{emulateapj}

\usepackage{float}
\usepackage{subfigure}
\usepackage{verbatim}
\usepackage{color}

\begin{document}

\title{Shell Collision Induced Kilo-Hertz Quasi-Periodic Oscillation in X-Ray Binaries}


\author
{Bi-Ping Gong$^{1}$, Ya-Ping Li$^{2}$}
\affil{$^{1}$School of Physics, Huazhong University of Science and Technology,
 Wuhan 430074, China}
\affil{$^{2}$Department of Astronomy and Institute of Theoretical Physics and Astrophysics,Xiamen University, Xiamen, Fujian 361005, China}






\begin{abstract}
{
Twin kilo-Hertz Quasi-periodic oscillation with ratio 3/2 has been found in some compact sources, which is believed to be related with  their innermost regions of accretion disks, and hence carrying  information of  gravity in strong regime.
However, more complicated phenomena have been revealed,
e.g.,  the twin kilo-Hertz Quasi-periodic oscillation of 4U 1820-20 start increasing at certain spectral state and then saturate upon reaching certain level.
Moreover, such quasi-periodic oscillation is not uniform and has multiple peaks,
which displays random feature. 
This letter  suggests that these  challenges could be easy to understand if the  quasi-periodic oscillation originates in jets of compact objects.
With a  seed periodicity originating in either  neutron star spin or accretion,  shell collision develops  in  jet.
The corresponding   twin kilo-Hertz quasi-periodic oscillations   automatically carry random feature, vary in frequency; and  couple with luminosity, spectral properties , which well account for the observation of 4U 1820-30.
New  quasi-periodic oscillation of 4U 1820-30 is predicted, which can  test the validity of this model. And the scenario is applicable   to other compact object like AGN. 
}
\end{abstract}

\section{Introduction}

Observations starting in 1996 with the Rossi X-ray Timing Explorer could detect faster variability, and it was found that neutron stars(NSs) and black holes(BHs) emit X-rays displaying quasi-periodic oscillation (QPO) with frequencies up to 1000 Hertz(Hz) or so,  which is believed to be related with  the accretion flow fre
around NSs and BHs very close to the compact object and hence can test general relativistic theory in strong gravity regime. However, more complicated phenomena revealed in some X-ray binary systems led to no consensus on the origin of these
high frequency QPOs, nor on what physical parameters determine these QPOs.

The source 4U 1820-30 is an atoll low mass X-ray binary(LMXB) in the globular cluster NGC 6624.
Two simultaneous kHz QPOs were detected in the persistent flux from 4U1820-30\citep{Zhang98}.
The source 4U 1820-30 undergoes a regular 176 day accretion cycle\citep{PT84},
which is further  reported  to be 172.780 days by \cite{Simon03}.

The source switches between low- and high-luminosity states, which differ by a factor of 3 in their X-ray luminosity.  In  spectral color-color  diagram  diagram (CCD), this corresponds to the evolution from the the island, lower banana, to upper banana branches.
As count rate increases from 1600 to 3200 count/s and above, the QPO frequency increases linearly with count rate below 2500counts/s and then saturated  while the flux of the source is still increasing\citep{Zhang98}.



There is a gap in the distribution of of QPO low frequencies between 600 and 650Hz\citep{Belloni05,Barret08}.
Moreover, the distribution of twin kHz QPOs are not uniform and have multiple peaks,
which display random feature\citep{Belloni05}.


Various theoretical models have been proposed to account for kHz QPO signals\citep{Miller98,Lamb01,Abra03,Klis97}. Although each model has merit in some aspects. They are still challenged by following facts:
(a) evolution and saturation of QPO frequency\citep{Zhang98}; (b)the random feature displaying as multiple peaks of QPOs\citep{Belloni05};
(c) correlation of such an evolving QPO with CCD; and (d) with the circle of long-term activity of a source, like 176 days of 4U1820-30.

This letter directly confronts these challenges.
With a seed periodicity at the central engine,  shell collision occurs in  jet, which can be treated by the technique applied in GRB. 

Consequently, the power density spectrum (PDS) response of  such a collision  can be a twin kHz QPOs
expected by observations.
Furthermore, the  evolution of such a twin QPOs automatically correlates with luminosity and spectral properties of  the source.
The model is described in Section 2， which is applied  to 4U1820-30  in Section 3, and corresponding prediction is made in Section 4.  

\section{Shell collision in jet}
The jets of NS  can behave much like jets from black holes. Typically,   collimated matter is blasted into space like water from a fire hose,  as jets of  Circinus X-1 measured by Chandra.
 When the jets reach the gas of interstellar space, they seem to evolve to radio-emitting gas, which apparently carry huge amounts of kinetic energy, releasing through jet-matter interaction.

Matter falling onto the surface of an NS is thought to be along closed magnetic field lines contacting the accretion disk. However, the corresponding geometry of outflow is not well understood. Owing to the remarkable similarity between jets of BH and NS, it is possible that the outflow is also launched through open field line like the Blandford-Znajek effect\citep{BZ77} in BH X-ray binaries.
In such case, the spin of  a NS affects the energy ejection at the line of sight (LOS), where the emission originated in shell collision is observed. 

For an accreting NS system with opening angle of jet of $1/\Gamma\sim 10^{\circ}$, a
misalignment angle between the jet axis and the spin axis, around the same level, $\sim 10^{\circ}$,
will not produce the  pulsation like a pulsar, since the emission beam is never completely out of LOS. Obviously such a spin of NS results in the episodic ejection at LOS even if the jet is orinially continuous.

Episodic ejection can also be reproduced  by  fluctuated accretion due to the interaction between the orbital motion of the flow and the NS rotation\citep{Miller98,Lamb01}, which predicts a
beat frequency, $\nu_b=\nu_{K}-\nu_{s}$  (where $\nu_{K}$ and $\nu_{s}$  are Keplerian frequency and spin frequency of NS respectively) responsible for high frequency QPO. In such case, the jet of such  NS is not relevant with its pulsation, as displayed in the pulsar nebular of Crab.


Either of the two processes discussed above, can work as seed periodicity, so that  shell collision can be generated in jet. 
In fact, the property of shell collision  has be studied extensively in GRB  prompt emission, which can be used in the case of QPO.
Assuming a periodicity in the central engine, ejectors are launched out, in which a rapid shell (subscript $r$) catches up a slower one (s), and merge to a single one ($m$).
The system behaves like an inelastic collision between two masses. By the conservation of energy and momentum, the Lorentz factor of the merged shell can be  calculated\citep{Kob97, Granot12}.

The evolution of the outflow can be studied by assuming that shells merge after collisions, which is approximately valid when all the internal energy generated by a collision of two shells is  radiatively efficient\citep{Zou06,Li08}. As the flow radius increases, the typical number of initial shells that merge into one single shell increases, so that the variance of the Lorentz factors of the resulting shells decreases.
In the case of  numerous shells (much larger than that of Li $\&$ Waxman\citep{Li08}), the collision scenario is that fast shells with approximately same speed collide with a  single shell which is relatively slow in speed and large in thickness and mass.

In the rest frame of the NS, the  time scale of $i$-th collision between  the rapid and merged shell (with dimensionless speeds $\beta_r$ and $\beta_m$ respectively) is given,
\begin{equation}\label{t}
T_i=\frac{\Delta_{tot}}{c(\beta_r-\beta_m)}
\end{equation}
where $\Delta_{tot}=\Delta_{gap}+\Delta_{sh}$, and
 $\Delta_{gap}/c$ and $\Delta_{sh}/c$ correspond to the period during which the central engine is off and on respectively.





Such an off and on can be realized by  the  flip  jet axis.
Suppose a shell is ejected between  $t_0$ and $t_1=t_0+\Delta_{sh}/c$, at direction $A$;
and between  $t_1$ and $t_2=t_1+\Delta_{tot}/c$, another shell is launched at direction $B$， which deviates from A for a few degrees ...
Consequently, for one circle of flip, there is one shell (ejector) launching out at time scale,  $T_1=(\Delta_{gap}+\Delta_{sh})/c$ (intrinsic time scale) at one direction.


In the co-moving frame of the rapid shell, the collision period of Eq.$\ref{t}$ becomes,
\begin{equation}\label{t1}
T_2=D_{-}T_i=\frac{D_{-}(\Delta_{tot})}{c(\beta_r-\beta_m)}
=\alpha~T_1
\end{equation}
where $\alpha=D_{-}/(\beta_r-\beta_m)$,  the value of which will be discussed later,
$D_{-}={1}/{\Gamma(1+\beta)}=\sqrt{\frac{1-\beta}{1+\beta}}$ is the Doppler factor relating the rest frame of NS and the rapid shell of Lorentz factor $\Gamma$.

Transforming the period, $T_2$, defined in the co-moving frame of the shell to observer's frame, we have,
\begin{equation}\label{tobs}
T_{obs,2}=T_2\Gamma(1-\beta\mu),
\end{equation}
where  $\mu=\cos\theta(t)$,  in which $\theta$ denotes the secular change of the misalignment angle between the jet axis and LOS.
Further more, $T_{obs,2}$ corresponds to an oscillation frequency of,  $\omega_2=2\pi/T_{obs,2}$.
As the jet precesses, the misalignment angle between the jet axis and LOS, $\theta$, varies,  so that the specific intensity  in comoving frame, $F_{\ast}$ , is Doppler boosted and measured in observer frame as,
\begin{equation}\label{Fobs}
 F_{obs}=D^3(t) F_{\ast}\exp(i\omega_2 t),
\end{equation}
where
$D(t)={1}/{\Gamma\{1-\beta\cos[\theta(t)+\delta\theta(t)]\}}$
 is the Doppler factor,
where $\delta\theta(t)=\delta\theta_0\exp(i\omega_1 t)$, with $\omega_1=2\pi/T_{1} $, represents rapidly flip of the jet axis relative to LOS.
Apparently, the shell collision induced L-C, as shown in Eq.$\ref{Fobs}$, couples not only with two oscillations, $\omega_1$ and $\omega_2$, but also  with the angle $\theta$. Thus,
the PDS response of Eq.$\ref{Fobs}$ at different  phase of jet precession (different $\theta$), gives different frequencies of QPOs.

Notice that if the shell collision is due to spin induced flip, $\omega_1=2\pi/T_{1} $, the oscillation varies with shell thickness, which increases with increasing core-site separation. And in the case of beat frequency induced periodicity in jet, a constant $\omega_1$ is expected, in which case,   $\delta\theta(t)$ corresponds to  fluctuated change of flux density given by Eq.$\ref{Fobs}$.

\section{Confronting with observation of 4U 1820-30}

\begin{figure}
\includegraphics[width=0.45\textwidth]{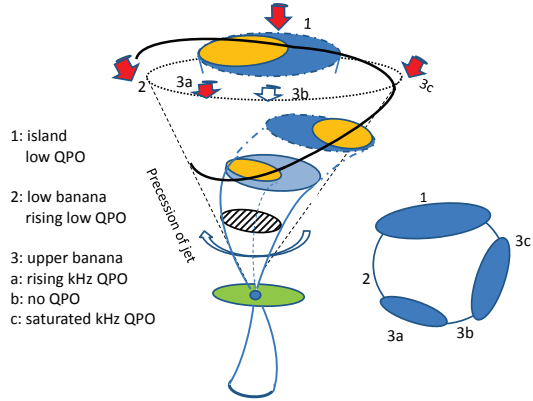}
\caption{\small Schematic plot of the precession geometry of LMXB 4U1820-30.  The opening angle of the jet is denoted by the small ellipse, and its range of flip is represented by the large ellipse. The jet precesses at a 176 days period, towards left, which is equivalent to the motion of LOS at opposite direction. The states of CCD correspond to different misalignment angle of the jet to  LOS.
\label{scenario}}
\end{figure}

The collision described by the above equations occurs in the  cone   represented by the big ellipses as shown in Fig.~$\ref{scenario}$, in which the instantaneous cone denoted by the small (yellow) ellipses   flip in.

Launched at time, $t_{0}$, the bulk speed of ejector $i$ is $c\beta_i$.
 The ejector-core
separation of site $i$ at time $t_{i}$ is $R_{\theta_{i}}=R_0+c\beta(t_{i}-t_{0})$. Owing to the jet precession, the  neighboring sites of ejector $i$ deviate from it both in direction and core-site separation.
Such a  continuous and  precessing jet  results in a spiral jet, as shown in Fig.~$\ref{scenario}$, in which the collision site of kHz QPO is part of the spiral, from 3a to 3c, as shown in Fig.~$\ref{scenario}$.


The jet precession  towards left is equivalent to the motion of LOS at opposite direction as shown in Fig.~$\ref{scenario}$. The onset of the shortest collision site corresponds to the start of kHz QPOs, which increase till saturation. After that, the increase of site-core separation,  $R_{\theta_{i}}$,  reduces the collision frequency, which is responsible for the QPO frequency of a few Hz.



%


As shown in Fig.~$\ref{scenario}$, the onset of  sequence, 1,2,3 and 4 is observed due to the clockwise precession of jet. This explains the over all state change of 4U1820-30,
the circle of island, low banana, upper banana of 4U1820-30.
And  the 176 days  is apparently the precession period  of jet as shown in Fig.~$\ref{scenario}$.

Such a preession period can be satisfied by the long term effect of a binary system. E.g.,  the precession speed predicted by the geodetic precession\citep{BO75} , is given by 
\begin{equation}\label{SL}
\Omega=\frac{1}{2}(\frac{GM_\odot}{c^3})^{2/3}(\frac{2\pi}{P_{b}})^{5/3}m_c(4m_p + 3m_c)/(m_c+m_p)^{4/3}/(1 - e^2)
\end{equation}
With orbital eccentricity $e=0.65$,  the orbital period $P_b=11$ minute(685s)\citep{Stella87},  and  companion mass, $m_c=0.06$\citep{Rapp87}(assuming $m_p=1.4$),
The precession period $P_p=2\pi/\Omega\approx 176$days can be satisfied. 

Same precession period can be given by the Newtonian-driving precession\citep{Larwood98} as well,
\begin{equation}\label{TQ}
P_p=\{\frac{3m_c}{7m_p}(\frac{r_d}{d})^{3/2}(\frac{m_p}{m_p + m_c})^{1/2}\cos\theta\}^{-1}P_b
\end{equation}
with disk radius $r_d=700r_s$ and the misalignment angle between accretion disk and orbital plane $\theta=\frac{\pi}{4}$, where $r_s$ is Schwarzschild radius and $d$ is the orbital separation.

At different $\theta$, e.g., $\theta$=0.16rad, 0.27rad, 0.35rad, we can have corresponding  PDS of the time series $F_{obs}$ determined by Eq.$\ref{Fobs}$. This actually predicts the evolution of QPO frequencies versus $\theta$.
And due to the change of $\theta$ varies the counts through the effect of Doppler boosting as shown by Eq.$\ref{Fobs}$, the correlation of counts versus $\theta$ is a natural result.

Therefore,  the evolution of the twin QPO frequencies versus the change of  flux predicted  by this model can be given by  Eq.$\ref{Fobs}$, which can be compared with observation\citep{Zhang98} as shown in Fig. $\ref{QPO2fre}$.
The fitting parameters are given in Table $\ref{parameters}$, in which jet opening angle is relatively large and opening angle precession cone is not much larger than that of the jet and the amplitude of flip of the jet. Therefore, its emission never goes down below observation threshold. Under such a configuration,  QPO can exist for the majority part of the precession period for this source, which well account for the observation\citep{Zhang98}.

The PDS of Eq.$\ref{Fobs}$ at different $\theta$ is  shown in Fig. $\ref{QPO2fre}$. This numerical result indicates that  the lower QPO  corresponds to the frequency of shell collision, $\omega_2/2\pi $; and the upper one corresponds to the coupling between the frequency of shell collision and the intrinsic frequency $\omega_2/2\pi + 1/T_1$.

The frequency discrepancy between the twin kHz QPOs is  $\nu_{u}-\nu_{L}=1/T_1\simeq$ 275Hz, in which  $T_1$ is a parameter related with the seed periodicity of the NS system.
The frequency ratio between the twin QPOs is, $\nu_{u}/\nu_{L}=1/(T_1\nu_{L})+1$, which can decrease from 1.6 to 1.3 with the increase of $\nu_{L}$ from 450Hz to 800Hz. This is  consistent with observations of 4U 1820-30\citep{Barret08}.

This scenario explains not only the increase of QPO frequencies\citep{Zhang98}, but also its correlation with change of flux.
As shown in Fig.~$\ref{scenario}$, when the slashed ellipse and ellipses near the sign 3a is precessing close to LOS, the first  collision site of kHz QPO (with shortest core-site separation) is onset. This actually corresponds to a decrease of  $\theta$, and accordingly an increase of $\omega_2$  as given by Eq.$\ref{tobs}$.
Until its minimum $\theta_{min}$ is reached between 3b and 3c of Fig.~$\ref{scenario}$.

Later on, $\theta$, the misalignment angle of the first collision site and LOS, continues increase,  so that LOS is aligning with collision site of larger core-site separation as denoted by  3c in Fig.~$\ref{scenario}$.



In such case, the collision parameters of Eq.$\ref{t}$-Eq.$\ref{Fobs}$ are unchange, and hence the
kHz QPO frequency is  unchanged. During this process, the soft emission from the disc contributes to the flux, which explains the saturation of QPO frequency with increase of flux\citep{Zhang98}.



A gap around the point (600,900)Hz in correlation line that links the upper QPO frequency to lower QPO frequency is reported\citep{Barret08}, in which  twin kHz QPOs are absent.

This process corresponds to the LOS denoted by  3b in Fig.~$\ref{scenario}$, in which the first collision site and its adjacent sites (of larger core-site separation) are aligned with LOS simultaneously (recall large opening angle of the jet). The superposition of the time series of such multi-site may result in noise feature in  PDS, which explains the QPO gap\citep{Barret08,Belloni05}.



In this new scenario,
random process is involved both  in splitting the continuous jet into discrete shells (ejectors),  and in shell collision process. Moreover, the clump in the accretion of NS systems  also contaminates the signal randomly.

On the other hand, with parameters of Table~1 and a speed discrepancy of $\beta_r - \beta_m \approx 17\%\sim6\%$, the parameter $\alpha$ is confined approximately between 1$\sim$3.
If $\alpha$ deviates from 1$\sim$3 significantly, the QPO frequency would be much lower, or much higher than kHz respectively, which  can  be both ruled out for the case of 4U 1820-30.


All these factors contribute to the multiple peaks of kHz QPO, which is consistent with the overall distribution of peaks of  QPO\citep{Belloni05}.
In fact, the new model predicts that multiple peaks of kHz QPO exist during the whole evolution of kHz QPOs, which can be tested by further data analysis.



The frequency discrepancy in some LMXB systems (e.g., Sco X-1) becomes flat at higher frequency range\citep{Lin11}. This can be naturally explained by the increase of the shell thickness, $\Delta_{sh}$, which increases $T_1$, and reduces  the frequency discrepancy between the twin QPOs. At the same time, QPO frequencies increase owing to  the decease of $\theta$.
Consequently, the change of frequency discrepancy  of Sco X-1\citep{Lin11} can be   interpreted  as  the evolution  $T_1$, as shown in Fig.~\ref{dgap}.

The  QPO frequency of 7Hz observed in 4U 1820-30\citep{Belloni04}, can also be interpreted by the increase of core-site separation and hence increase of the shell thickness, $\Delta_{sh}$, and $T_1$, and thus low QPO frequency. Obviously, such low oscillation frequency can occur at state after state 3c and before state 1 as shown in Fig.~$\ref{scenario}$.

As a measure of the signal strength, the root-mean squared  (RMS) amplitude   is proportional to the square root of the peak power contribution to the PDS. The evolution of RMS at different $\theta$ can be obtained through Eq.$\ref{Fobs}$ either. As shown in Fig. $\ref{rms}$,
the trend of the evolution of RMS of the lower QPO frequency  is consistent with the observation\citep{TOROK09}. Although the absolute amplitude of RMS in the numerical calculation is much larger than that of observation, owing to the signal to noise ratio of simulated signal is much stronger than that of true data.

\section{Discussion and prediction}
\begin{figure}
\includegraphics[width=0.5\textwidth]{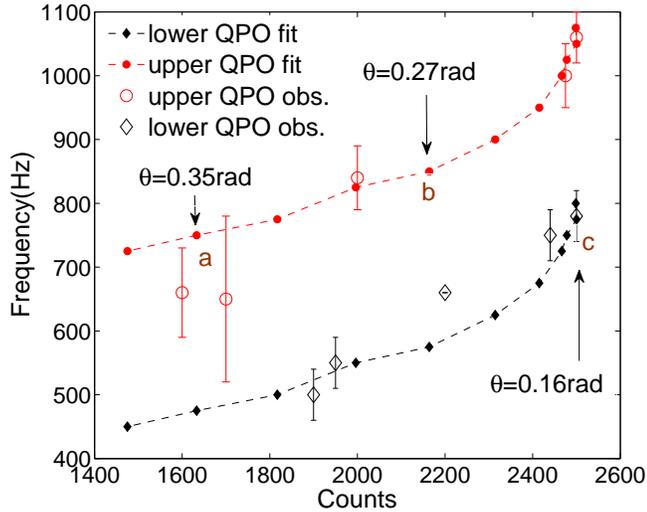}
\caption{\small  Observed and fitted QPO centroid frequencies versus count rate.  Observational data is from \cite{Zhang98}. Different  misalignment angle between LOS and jet axis, $\theta$, determines twin kHz QPO frequency and counts through Eq.~$\ref{Fobs}$. The time elapsed in the observer's frame at point $b$ and point $c$ from $a$ are 16.3days and 37.0days, respectively.
\label{QPO2fre}}
\end{figure}

\begin{figure*}[]
\centering
\subfigure[]{\label{rms}
\includegraphics[ width=0.33\textwidth]{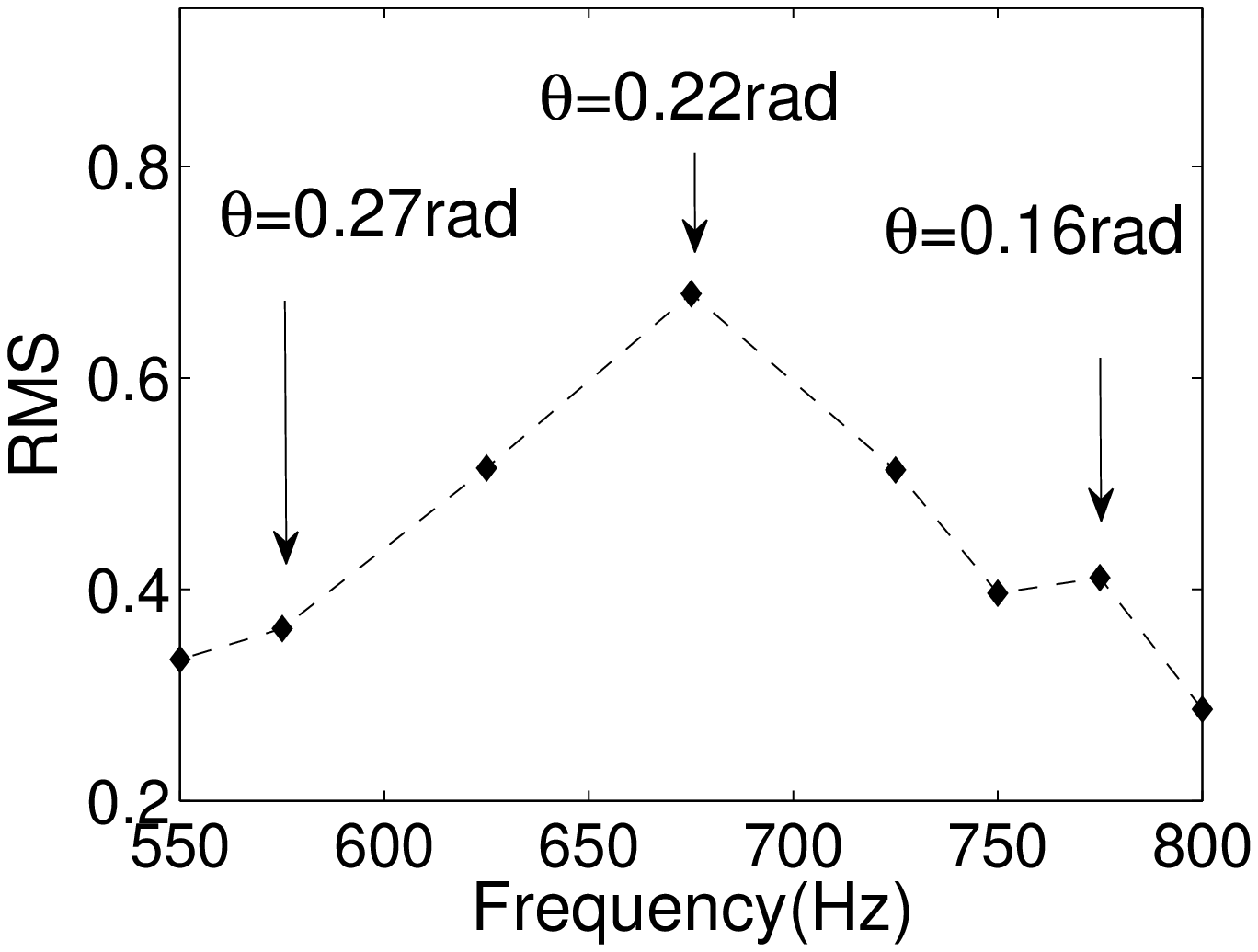}}
\hspace{-0.5cm}
\subfigure[]{\label{pds}
\includegraphics[ width=0.33\textwidth]{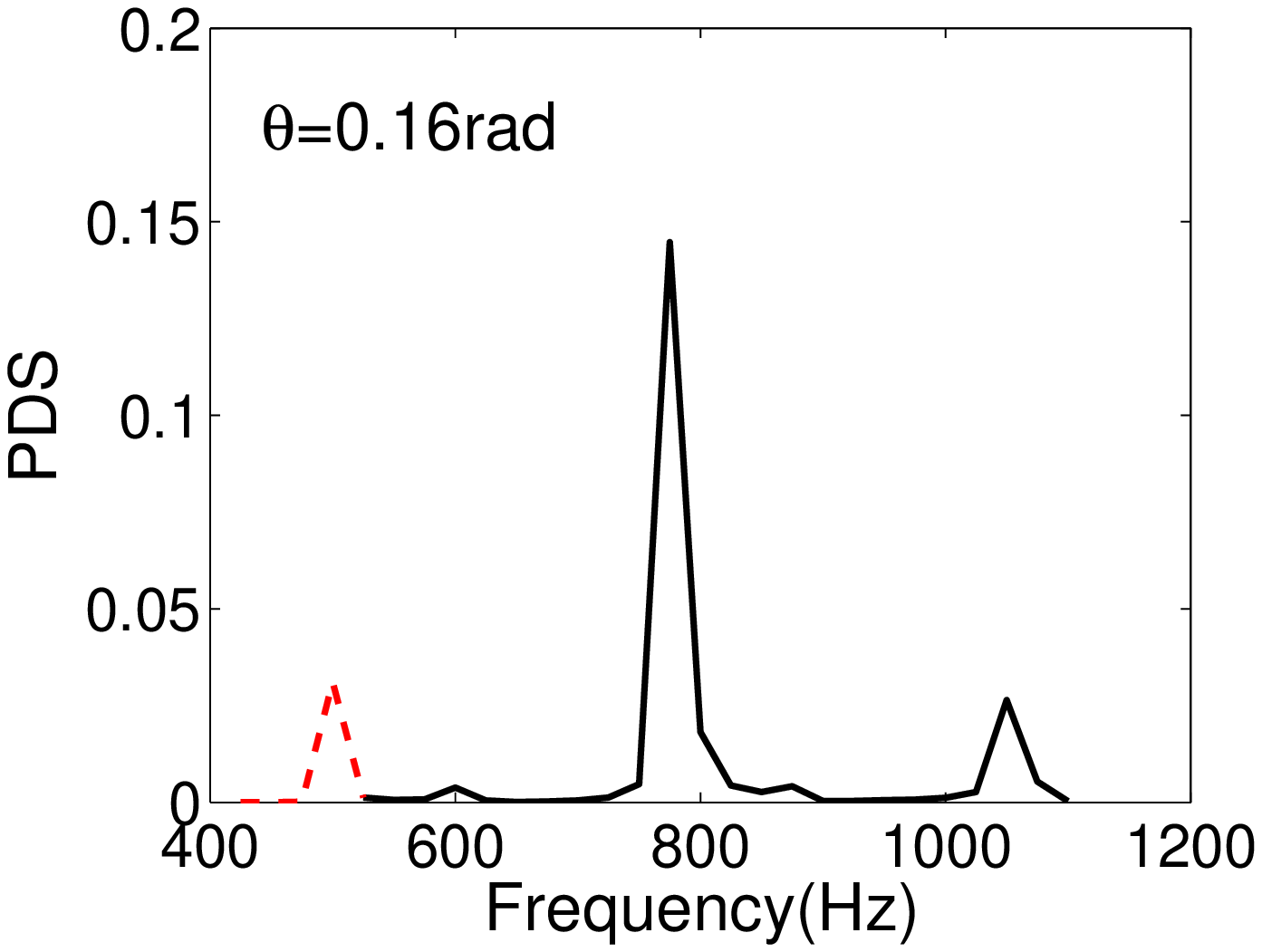}}
\hspace{-0.5cm}
\subfigure[]{\label{dgap}
\includegraphics[ width=0.33\textwidth]{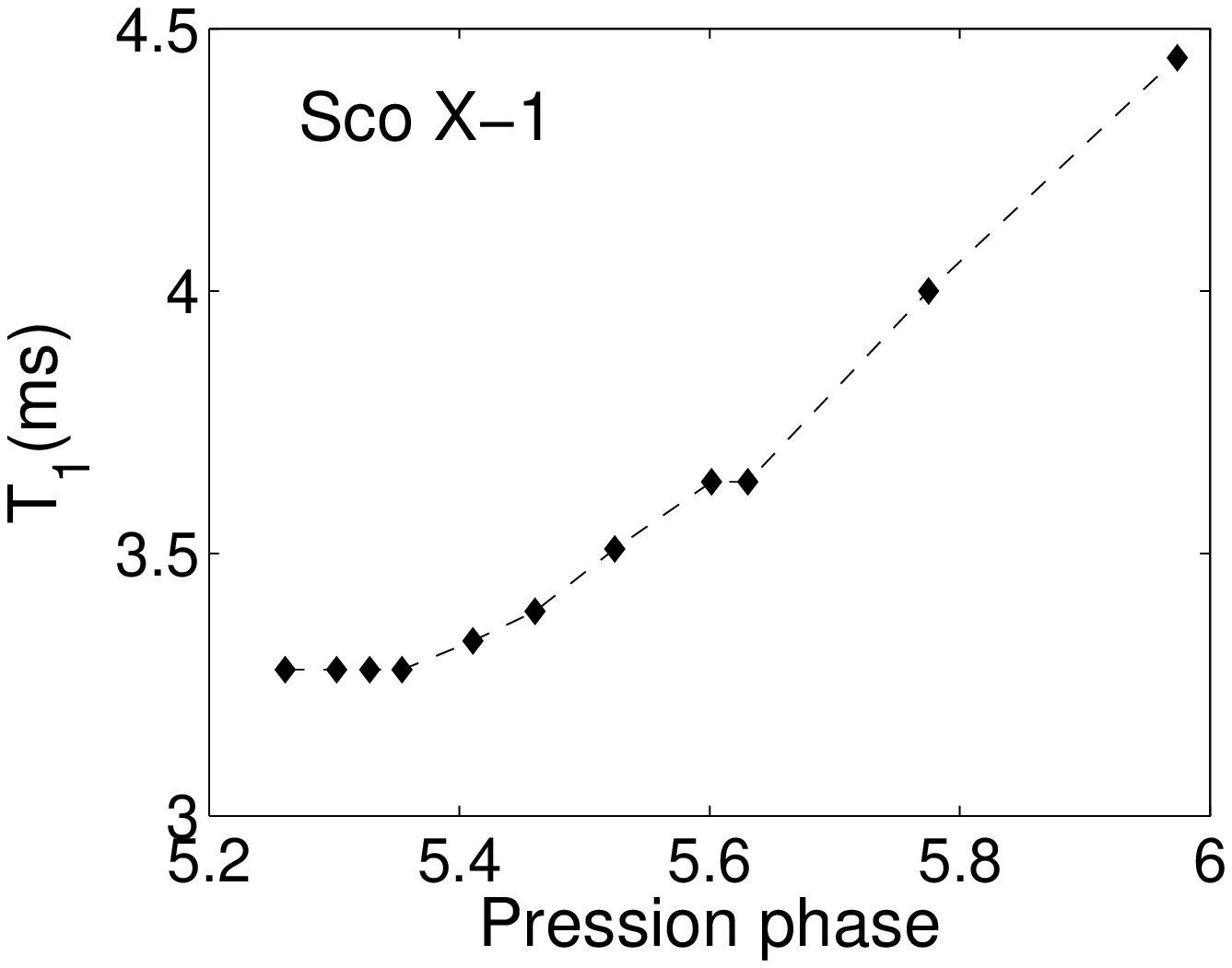}}
\caption{\small(a). Fitted RMS versus lower QPO frequency consisting with the observed trend of 4U 1820-30\citep{TOROK09}. The time elapsed between $\theta=0.27$ and $\theta=0.22$ is 9.0days. (b). Power Density Spectrum of 4U 1820-30 for a specific precession state with   $\theta$=0.16rad. The predicated QPO frequency is indicated by the red dashed peak. (c). The evolution of $T_1$ with precession phase responsible for the frequency relationship of Sco X-1\citep{Lin11}.}\label{fig:sub}
\end{figure*}

In such a simple scenario, with six parameters as shown in Table $\ref{parameters}$, the evolution of the twin kHz QPO versus counts is  simulated as shown in   Fig. $\ref{QPO2fre}$, which explains the correlation of  the twin kHz QPOs with the luminosity, and hence the spectral state of  4U 1820-30.
Simultaneously,  it also explains  the RMS trend of the lower QPO, as shown  Fig. $\ref{rms}$. The fitting of  Fig. $\ref{rms}$ and Fig. $\ref{QPO2fre}$ together suggests that  the observational feature of twin QPOs can be reproduced simply by an emission site suffering  Doppler effects, which in turn suggests that kHz QPO may originated in shell collision in jet of  X-ray binaries.

In this model, the state of a source with jet precession can be represented by the misalignment angle between the jet axis and LOS,  e.g., $\theta=0.16$rad, which corresponds to the parameter, $S_a$, measuring the position of an atoll source within its CCD\citep{HK89}.
As a result, the model can be tested by the  time elapsed between states (denoted by different $\theta$, and hence different $S_a$), as predicted by $\ref{QPO2fre}$ and  Fig. $\ref{rms}$.

The simulation of PDS of  Eq.$\ref{Fobs}$ also exhibits  a QPO  with frequency,  $\nu = \nu_{L}-$ 275Hz, as denoted by dashed peak in Fig.~\ref{pds}.
With a lower RMS amplitude, the predicated frequency can vary at the range, 325$\sim$ 525Hz, as $\nu_{L}$ changes at the range of 600$\sim$800Hz.
The measurement  of such predicted  QPO frequency will be direct test of the model.

\acknowledgments
We thank B. Zhang, Z. Li, Y.C. Zou and W.H. Lei  for  helpful
discussions. This research is supported by the
National Natural Science Foundation of China, under grand
NSFC11178011.

{}


\begin{table}
  \begin{center}
    \caption{ The fitting parameters for the QPO model}\label{parameters}
 \begin{tabular}{cccccc}
 \\
  \hline \hline
  $\lambda$(rad) & $i$(rad) & $T_1$(ms) & $\Gamma$ & $\delta\theta_0$(rad) &$\alpha$\\
  0.29 & 0.14 & 3.6 & 3.0 & 0.3 & 1.7\\
  \hline \hline

 \end{tabular}
\end{center}
{\small  NOTE: $\lambda$ is the opening angle of  jet precession, $i$ is the misalignment angle between jet rotation axis and LOS, $\Gamma$ is the Lorentz factor of jet and $\delta\theta_0$ is flip amplitude of jet. }
\end{table}

\end{document}